\documentclass[preprint,aps,showpacs]{revtex4}

\usepackage{mathrsfs}
\usepackage{amsmath}
\usepackage{amssymb}
\usepackage{epsfig}
\usepackage{graphicx}
\usepackage{booktabs}
\usepackage{array}
\usepackage{multirow}

\begin{document}

\renewcommand{\arraystretch}{0.5}
\newcommand{\beq}{\begin{eqnarray}}
\newcommand{\eeq}{\end{eqnarray}}
\newcommand{\non}{\nonumber\\ }

\title{Time-dependent $CP$-violations of $B(B_s)$ decays in the perturbative QCD approach}
\author{Xin Yu$^a$}
\author{Zhi-Tian Zou$^{a,b}$}
\author{Cai-Dian L\"{u}$^a$}\email{lucd@ihep.ac.cn}
\affiliation{
a. Institute  of  High  Energy  Physics  and  Theoretical  Physics Center for Science Facilities,
Chinese Academy of Sciences, Beijing 100049, China \\
b. Department of Physics, Yantai University, Yantai 264005, China }
\date{\today}


\begin{abstract}

We study the decay modes of $B_{s}^{0}(\bar{B}_{s}^{0})\rightarrow D_{s}^{\pm} K^{\mp} $, $B_{s}^{0}(\bar{B}_{s}^{0})\rightarrow D^{\pm} \pi^{\mp} $  and $B^{0}(\bar{B}^{0})\rightarrow D^{\pm} \pi^{\mp} $ in the perterbative QCD approach based on $k_T$ factorization, including the branching ratios and $CP$ violation parameters which provide a clear way to extract the Cabibbo-Kobayashi-Maskawa angle $\gamma$. Our results of branching ratios of $B_{s}^{0}(\bar{B}_{s}^{0})\rightarrow D_{s}^{\pm} K^{\mp} $ and $B^{0}(\bar{B}^{0})\rightarrow D^{\pm} \pi^{\mp} $ and the CP asymmetry of $B^{0}(\bar{B}^{0})\rightarrow D^{\pm} \pi^{\mp} $ agree well with the experimental data. We also give the predictions of the other observables, which provide some guidance for experiments in the future, especially for LHCb experiment.

\end{abstract}

\maketitle

\section{Introduction}

The study of the non-leptonic two body B meson decays plays an important role in extracting the Cabbibo-Kobayashi-Maskawa(CKM) matrix elements, whose phase provides the source of $CP$ violation in the Standard Model. Unlike the precise measurements of angle $\beta$ and $\alpha$, the experimental uncertainty of the CKM angle $\gamma$ is large, roughly about $10^{\circ}$\cite{Beringer:1900zz}. To extract the angle $\gamma$ precisely is one of the major goals no matter in today's LHCb experiment or in the future SuperB factory experiment. The most popular way to measure $\gamma$ is through decays $B^{\pm} \rightarrow DK^{\pm}$. There are three well-established methods, including the Gronau-London-Wyler method\cite{Gronau:1990ra,Gronau:1991dp,Dunietz:1991yd}, the Atwood-Dunietz-Soni method\cite{Atwood:1996ci,Atwood:2000ck}, and the Giri-Grossman-Soffer-Zupan method\cite{Giri:2003ty}, with different final states of D meson decays respectively.

Time-dependent studies of non-$CP$ eigenstates provide another method to extract the CKM angle $\gamma$. This method was first proposed in \cite{Aleksan:1991nh} for $B_{s}^{0}(\bar{B}_{s}^{0})\rightarrow D_{s}^{\pm} K^{\mp} $ and $B^{0}(\bar{B}^{0})\rightarrow D^{\pm} \pi^{\mp} $ decays and then was further studied by \cite{Fleischer:2003yb,DeBruyn:2012jp}. A similar decay mode $B_{s}^{0}(\bar{B}_{s}^{0})\rightarrow D^{\pm} \pi^{\mp} $ used to determine $\gamma$ is proposed in \cite{Hong:2005vq,Li:2003wg}. All these three kinds of channels share the same character that both a pure $B^0_{(s)}$ and its antiparticle $\bar{B}^0_{(s)}$ can decay to the same final states, which leads to the $CP$ violation in the interference between decays with and without mixing. There are difficulties in experiment to measure these decays because it needs a large sample of $B^0_{(s)}$ decays and one has to distinguish the rapid $B^0_{(s)}-\bar{B}^0_{(s)}$ oscillations. However recently LHCb has performed the first time-dependent analysis of $B_{s}^{0}(\bar{B}_{s}^{0})\rightarrow D_{s}^{\pm} K^{\mp} $ decays\cite{Blusk:2012v3}. In this paper we explore this kind of time-dependent $CP$ violations of the above three decay modes, hoping to provide some guidance for experiments in the future. The method we use is the  purturbative QCD(PQCD) approach, based on $k_T$ factorization, which is successfully applied in the hadronic two body decays of B mesons, especially for estimation of the direct $CP$ asymmetries\cite{Hong:2005wj}.

This paper is organized as follows: In Sec.\ref{sec:forma}, we make a brief introduction of time-dependent $CP$-violations and explain how the CKM angle $\gamma$ can be extracted. Then we present the formalism and wave functions used in the PQCD approach in Sec.\ref{sec:pqcd}. The numerical results of branching ratios and $CP$ violation parameters and phenomenological discussions are given in Sec.\ref{sec:result}. Finally, Sec.\ref{sec:sum} is a short summary.

\section{Time-dependent $CP$-violations}\label{sec:forma}

In the neutral $B_q^0-\bar{B_q^0}$($q=d,s$) mixing system, the light(L) and heavy(H) mass eigenstates are related to the flavor eighenstates $B_q^0$ and $\bar{B_q^0}$ by
\begin{subequations}
\begin{align}
|B_L\rangle &= p|B_q^0\rangle + q|\bar{B_q^0}\rangle , \\
|B_H\rangle &= p|B_q^0\rangle - q|\bar{B_q^0}\rangle .
\end{align}
\end{subequations}
Here we define the mass difference $\Delta m_q = m_H - m_L$, the total decay width difference $\Delta \Gamma_q = \Gamma_L -\Gamma_H$ and the average decay width $\Gamma_q = (\Gamma_L +\Gamma_H)/2$. In the Standard Model, $q/p$ is given by
\begin{eqnarray}
\frac{q}{p} \approx \frac{V_{tq}V_{tb}^*}{V_{tq}^*V_{tb}},
\end{eqnarray}
as a ratio of CKM elements, with $|q/p| \approx 1$. For $B_d$ system, $(\frac{q}{p})_{B_d} \approx e^{-2i\beta}$ with $\beta=\arg(-\frac{V_{cd}V_{cb}^*}{V_{td}V_{tb}^*})$ denoting the CKM angle. Similarly, for $B_s$ system, we have $(\frac{q}{p})_{B_s} \approx e^{-2i\beta_s}$. In order to get the same formulae with $B_d$, we define $\beta_s = \arg(-\frac{V_{cs}V_{cb}^*}{V_{ts}V_{tb}^*})$, with an additional minus sign compared with the definition of Heavy Flavor Averaging Group\cite{Amhis:2012bh}. From the definition we can see that $\beta_s$ is very small with the experimental average $2\beta_s=(-2.5_{-4.9}^{+5.2})^{\circ}$\cite{Amhis:2012bh}.

The time-dependent decay rates of $B_q^0$ and $\bar{B_q^0}$ decays to a final state $f$ are given by\cite{Amhis:2012bh}
\begin{subequations}
\begin{align}
\lefteqn{\Gamma(B_q^0(t)\rightarrow f) = \frac{1}{2} N_f |A_f|^2 e^{-\Gamma_q t} (1+|\lambda_f|^2)} \nonumber\\
&& \times \lbrace \cosh(\frac{\Delta\Gamma_q t}{2})+A_f^{\Delta\Gamma}\sinh(\frac{\Delta\Gamma_q t}{2})
+C_f\cos(\Delta m_q t)-S_f\sin(\Delta m_q t)\rbrace , \\
\lefteqn{\Gamma(\bar{B_q^0}(t)\rightarrow f) = \frac{1}{2} N_f |A_f|^2 (\frac{p}{q})^2 e^{-\Gamma_q t} (1+|\lambda_f|^2)} \nonumber\\
&& \times \lbrace \cosh(\frac{\Delta\Gamma_q t}{2})+A_f^{\Delta\Gamma}\sinh(\frac{\Delta\Gamma_q t}{2})
-C_f\cos(\Delta m_q t)+S_f\sin(\Delta m_q t)\rbrace ,
\end{align}
\end{subequations}
where $N_f$ is the normalization factor and $A_f$ is the decay amplitude of $B_q^0 \rightarrow f$. In the $B_d$ system $\Delta\Gamma_d$ could be neglectable, while in the $B_s$ system $\Delta\Gamma_s$ is quite important since $\Delta\Gamma_s / \Gamma_s = +0.144\pm0.021$\cite{Amhis:2012bh}.
The definition of $\lambda_f$ is
\begin{eqnarray}
\lambda_f = \frac{q}{p} \frac{A(\bar{B_q^0}\rightarrow f)}{A(B_q^0\rightarrow f)}.
\label{equ:lamdaf}
\end{eqnarray}
$\lambda_f$ can also be expressed as a complex number with a phase $\Delta-(\gamma+2\beta_q)$, where $\Delta$ is the difference of strong interaction final-state phase.

The $CP$ violation parameters are expressed as\cite{Amhis:2012bh}
\begin{eqnarray}
C_f = \frac{1-|\lambda_f|^2}{1+|\lambda_f|^2} , \qquad
S_f =\frac{2Im(\lambda_f)}{1+|\lambda_f|^2} ,  \qquad
A_f^{\Delta\Gamma} =-\frac{2Re(\lambda_f)}{1+|\lambda_f|^2} .
\label{equ:cpob}
\end{eqnarray}
If the final state is the $CP$-conjugate state $\bar{f}$, we have observables such as $\lambda_{\bar{f}}$, $C_{\bar{f}}$, $S_{\bar{f}}$ and $A_{\bar{f}}^{\Delta\Gamma}$, which have similar formulae with the replacement $f\rightarrow\bar{f}$ in Eqs.(\ref{equ:lamdaf}) and (\ref{equ:cpob}). Since there is only tree contribution in our considered decays, no direct $CP$ violation will appear here. We can get $|\lambda_f|=1/|\lambda_{\bar{f}}|$ and $C_{\bar{f}}=-C_f$.

Take $B_s$ decay for example, from the definition above we have
\begin{eqnarray}
\lambda_f = |\lambda_f|e^{i(\Delta-(\gamma+2\beta_s))},  \qquad
\lambda_{\bar{f}} = \frac{1}{|\lambda_f|}e^{i(-\Delta-(\gamma+2\beta_s))}.
\label{equ:twol}
\end{eqnarray}
Then the weak phase $(\gamma+2\beta_s)$ can be determined by
\begin{eqnarray}
\lambda_f \cdot \lambda_{\bar{f}} = e^{-2i(\gamma+2\beta_s)}.
\end{eqnarray}
Here $\lambda_f$ and $\lambda_{\bar{f}}$ are experimental observables and $\beta_s$ can be measured separately, therefore the CKM angle $\gamma$ can be extracted.

\section{Perturbative calculation in the PQCD approach}\label{sec:pqcd}

The weak decays considered in this paper are belong to the type of $B \rightarrow D(\bar{D})P$ decays, where $P$ denotes a pseudoscalar meson. There are only tree operators contributing to the weak effective Hamiltonian, which means that there is no penguin pollution. For the $B \rightarrow DP$ decays, the Hamiltonian can be written as
\begin{eqnarray}
H_{eff}=\frac{G_{F}}{\sqrt{2}}\,V_{ub}^{*}V_{cq}\left[C_{1}(\mu)O_{1}(\mu)\,+\,C_{2}(\mu)O_{2}(\mu)\right],
\end{eqnarray}
where $V_{ub}$ and $V_{cq}$ denote the CKM matrix elements with $q=d,s$ and $C_{1,2}(\mu)$ are Wilson coefficients at the renormalization scale $\mu$. The four-quark tree operators are
\begin{eqnarray}
O_{1}\,=\,(\bar{b}_{\alpha}u_{\beta})_{V-A}(\bar{c}_{\beta}q_{\alpha})_{V-A},
\;O_{2}\,=\,(\bar{b}_{\alpha}u_{\alpha})_{V-A}(\bar{c}_{\beta}q_{\beta})_{V-A},
\end{eqnarray}
with $(\bar{b}_{\alpha}u_{\beta})_{V-A}\,=\,\bar{b}_{\alpha}\gamma^{\mu}(1-\gamma^{5})u_{\beta}$. Here $\alpha$ and $\beta$ stand for color indices. Considering the $B \rightarrow \bar{D}P$ decays, the Hamiltonian is given by
\begin{eqnarray}
H_{eff}=\frac{G_{F}}{\sqrt{2}}\,V_{cb}^{*}V_{uq}\left[C_{1}(\mu)O_{1}(\mu)\,+\,C_{2}(\mu)O_{2}(\mu)\right],
\end{eqnarray}
with the tree operators
\begin{eqnarray}
O_{1}\,=\,(\bar{b}_{\alpha}c_{\beta})_{V-A}(\bar{u}_{\beta}q_{\alpha})_{V-A},
\;O_{2}\,=\,(\bar{b}_{\alpha}c_{\alpha})_{V-A}(\bar{u}_{\beta}q_{\beta})_{V-A}.
\end{eqnarray}

Dealing with hadronic B decays, one needs to prove factorization so that the perturbative QCD is applicable. Up to now, the factorization is only proved in the leading order of $1/m_B$ expansion\cite{Bauer:2001yt,Bauer:2002nz,Bauer:2004tj}. Working in this order, the light quarks in the final state mesons are in a collinear region. All the three quarks from b quark decay get large momentum, which are automatic collinear quarks. The spectator quark from B meson, which is soft, thus needs a hard gluon to transfer momentum. Finally in the PQCD approach, the decay amplitude can be factorized into the following form,
\begin{eqnarray}
\mathcal
{A}\;\sim\;&&\int\,dx_{1}dx_{2}dx_{3}b_{1}db_{1}b_{2}db_{2}b_{3}db_{3}\nonumber\\
&&\times
Tr\left[C(t)\Phi_{B}(x_{1},b_{1})\Phi_{M_{2}}(x_{2},b_{2})\Phi_{M_{3}}(x_{3},b_{3})
H(x_{i},b_{i},t)S_{t}(x_{i})e^{-S(t)}\right],
\end{eqnarray}
as a convolution of the Wilson coefficients $C(t)$, the hard scattering kernel $H(x_{i},b_{i},t)$, and the light-cone wave functions of mesons $\Phi_{M}(x,b)$. Here $x_i$ is the momentum fraction of the valence quark, $b_i$ is the conjugate variable of a quark's transverse momentum $k_{iT}$, and $t$ denotes the largest energy scale in the hard part $H(t)$. The jet function $S_{t}(x_{i})$ comes from the the threshold resummation that smears the end-point singularities on $x_i$. The Sudakov factor $e^{-S(t)}$, resulting from the resummation of double logarithm, suppresses the soft dynamics effectively so that the perturbative calculation of the hard part is applicable.

The light-cone wave functions of the initial and final state mesons describe the non-perturbative contributions that can not be calculated perturbatively. Fortunately they are universal for all decay modes, i.e. process-independent. The $B$ meson and $B_s$ meson share the same structure of wave function, but with different values of parameters due to a small $SU(3)$ breaking effect. The $B_q$ light-cone matrix element are always decomposed as\cite{Grozin:1996pq,Beneke:2000ry}
\begin{eqnarray}
&&\int d^4z e^{ik_1\cdot z}\langle0|\bar{b_\alpha}(0)|q_\beta(z)|B_q(P_1)\rangle \nonumber\\
&=&\frac{i}{\sqrt{6}}\{(\makebox[-1.5pt][l]{/}P_1+M_B)\gamma_5[\phi_{B_q}(k_1)
-\frac{\makebox[-1pt][l]{/}n-\makebox[-1pt][l]{/}v}{\sqrt{2}}\bar{\phi}_{B_q}(k_1)]\}_{\beta\alpha}.
\end{eqnarray}
Here $n$ and $v$ are dimensionless light-like unit vectors pointing to the plus and minus directions, respectively. From the above equation,we can see that there are two distribution amplitudes. However, we always neglect $\bar{\phi}_{B_q}(k_1)$ in our calculation because it gives numerically small contribution\cite{Lu:2002ny}. For the distribution amplitude $\phi_{B_q}$ in the b-space, we choose\cite{Lu:2002ny,Kurimoto:2001zj}
\begin{eqnarray}
\phi_{B_q}(x,b)=N_{B}x^{2}(1-x)^{2}\exp\left[-\frac{1}{2}
\left(\frac{m_{B}x}{\omega_{B}}\right)^{2}\,-\,\frac{\omega_{B}^{2}b^{2}}{2}\right],
\end{eqnarray}
with $N_B$ as the normalization constant. We choose the shape parameter $\omega_B = (0.4 \pm 0.04)$ GeV and the decay constant $f_B=(0.19\pm0.02)$ GeV for the $B$ meson. While for the $B_s$ meson, we take $\omega_{B_s} = (0.5 \pm 0.05)$ GeV and $f_{B_s}=(0.24\pm0.03)$ GeV.

For the $D$ meson, the light-cone distribution amplitude up to twist-3 are defined by \cite{Kurimoto:2002sb,Li:2008ts,Zou:2009zza,Li:2009xf}
\begin{eqnarray}
\langle D(P)|q_{\beta}(z)\bar{c}_{\alpha}(0)|0\rangle
\,&=&\,\frac{i}{\sqrt{6}}\int_{0}^{1}dx\,e^{ixP\cdot
z}\left[\gamma_{5}(\makebox[-1.5pt][l]{/}P\,+\,m_{D})\phi_{D}(x,b)\right]_{\alpha\beta}.
\end{eqnarray}
We choose the same form of the distribution amplitude $\phi_{D}$ as in Refs.\cite{Li:2008ts,Zou:2009zza,Li:2009xf}
\begin{eqnarray}
\phi_{D}(x,b)\,=\,\frac{1}{2\sqrt{2N_{c}}}\,f_{D}\,6x(1-x)\left[1+C_{D}(1-2x)\right]
\exp\left[\frac{-\omega^{2}b^{2}}{2}\right],
\end{eqnarray}
with $C_{D}=0.8\pm0.1$, $\omega=0.1$ GeV and $f_{D}=207$ MeV  for $D (\bar{D}$) meson
and $C_{D_s}=0.6\pm0.1$, $\omega_s=0.2$ GeV and $f_{D_{s}}=258$ MeV  for $D_{s} (\bar{D}_{s})$ meson.

For the light pseudoscalar meson, the light-cone distribution amplitude is given by
\begin{eqnarray}
&&\langle P(P) |\bar{q}_{2\beta}(z)q_{1\alpha}(0) |0\rangle \nonumber\\
&=& -\frac{i}{\sqrt{6}}\int_0^1 dx\,e^{ixP\cdot z}\gamma_{5}[\{\makebox[-1.5pt][l]{/}P\phi_{P}^{A}(x)+m_{0}
\phi_{P}^{P}(x)+m_{0}(\makebox[0pt][l]{/}n\makebox[0pt][l]{/}v-1)\phi_{P}^{T}(x)]_{\alpha\beta},
\end{eqnarray}
where $m_0$ as the chiral scale parameter is defined by $m_0=\frac{M_P^2}{m_{q_1}+m_{q_2}}$. The distribution amplitudes $\phi_{P}^{A}(x)$, $\phi_{P}^{P}(x)$ and $\phi_{P}^{T}(x)$ are usually expanded by the Gegenbauer polynomials, and their expressions can be found in Refs.\cite{Ball:1998tj,Ball:1998je,Ball:2004ye,Ball:2006wn}.

\begin{figure}[]
\begin{center}
\includegraphics[scale=0.8]{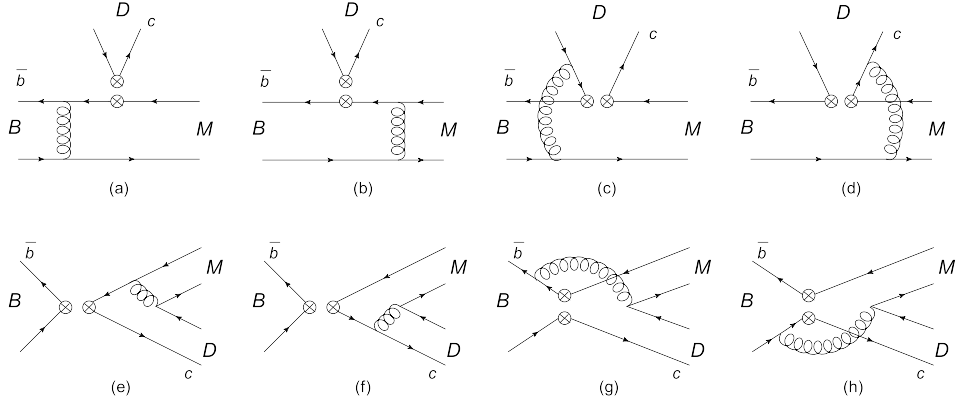}
\caption{Feynman diagrams contributing to the
$B\rightarrow DP$ decays in the PQCD approach. }
\label{fig:feyd}
\end{center}
\end{figure}

The hard part $H(t)$ is process dependent but perturbatively calculable. It involves the effective four-quark operators and the necessary hard gluon, which connects the four-quark operator with the spectator quark\cite{Lu:2000hj}. There are eight leading order diagrams that contribute to the $B \rightarrow DP$ decays, which are shown in Fig.\ref{fig:feyd}. For the $B \rightarrow \bar{D}P$ decays, there are similar diagrams but with different CKM matrix elements and different position of final state mesons. The first two diagrams in Fig.\ref{fig:feyd} are called factorizable emission diagrams, because their decay amplitudes can be factorized into the decay constant of the emitted meson and the transition form factor of B to another meson. The diagrams in Fig.\ref{fig:feyd}(c) and \ref{fig:feyd}(d) are nonfactorizable emission diagrams including the contributions of all three meson wave functions.  Fig.\ref{fig:feyd}(e) and \ref{fig:feyd}(f) stand for factorizable annihilation diagrams. For the last two diagrams in Fig.\ref{fig:feyd} called nonfactorizable annihilation diagrams, all three meson wave functions are involved in the decay amplitudes. The explicit expressions of all decay amplitudes for the above eight diagrams can be found in Refs.\cite{Li:2008ts,Zou:2009zza}.

\section{NUMERICAL RESULTS AND DISCUSSIONS}\label{sec:result}

\begin{table}[!h]
\centering
 \caption{Branching ratios calculated in the PQCD approach with experimental data\cite{Beringer:1900zz,Aaij:2012zz}.}
 \vspace{0.2cm}
\begin{tabular}[t]{l!{\;\;\;\;}c!{\;\;\;\;}c}
\hline\hline
 \multirow{2}{*}{Decay Modes}& \multirow{2}{*}{Br(theo)} & \multirow{2}{*}{Br(exp)}\\
 &&\\
 \hline
\vspace{0.2cm}
\multirow{1}{*}{$B_{s}^{0}\rightarrow D_{s}^{+} K^{-}$} &
\multirow{1}{*}{$(2.64^{+1.29+0.13+0.23}_{-0.96-0.18-0.21})\times 10^{-5}$} &
\multirow{2}[6]{*}{$(1.9\pm0.12\pm0.13^{+0.12}_{-0.14})\times 10^{-4}$} \\
\vspace{0.2cm}
\multirow{1}{*}{$B_{s}^{0}\rightarrow D_{s}^{-} K^{+}$} &
\multirow{1}{*}{$(1.57^{+0.88+0.42+0.09}_{-0.62-0.45-0.05})\times 10^{-4}$} & \\
\vspace{0.2cm}
\multirow{1}{*}{$B_{s}^{0}\rightarrow D^{+} \pi^{-} $ } &
\multirow{1}{*}{$(1.53^{+0.53+0.13+0.13}_{-0.45-0.15-0.12})\times 10^{-7}$} &
\multirow{1}{*}{$\ldots$} \\
\vspace{0.2cm}
\multirow{1}{*}{$B_{s}^{0}\rightarrow D^{-} \pi^{+} $ } &
\multirow{1}{*}{$(1.81^{+0.58+0.29+0.11}_{-0.50-0.34-0.05})\times 10^{-6}$} &
\multirow{1}{*}{$\ldots$}  \\
\vspace{0.2cm}
\multirow{1}{*}{$B^{0}\rightarrow D^{+} \pi^{-} $ } &
\multirow{1}{*}{$(8.21^{+3.51+0.45+0.77}_{-2.66-0.56-0.69})\times 10^{-7}$} &
\multirow{1}{*}{$(7.8\pm1.4)\times 10^{-7}$} \\
\vspace{0.4cm}
\multirow{1}{*}{$B^{0}\rightarrow D^{-} \pi^{+} $ } &
\multirow{1}{*}{$(2.42^{+1.22+0.38+0.13}_{-0.88-0.56-0.06})\times 10^{-3}$} &
\multirow{1}{*}{$(2.68\pm0.13)\times 10^{-3}$} \\
\hline\hline
\end{tabular}\label{t1}
\end{table}

\begin{table}[!h]
\centering
 \caption{$CP$ violation parameters calculated in the PQCD approach($f$=$D_s^+K^-$, $D^+\pi^-$).}
 \vspace{0.2cm}
\begin{tabular}[t]{c!{\;\;\;\;}c!{\;\;\;\;}c!{\;\;\;\;}c}
\hline\hline
 \multirow{2}{*}{Decay Modes} &
 \multirow{2}{*}{$B_{s}^{0}(\bar{B}_{s}^{0})\rightarrow D_{s}^{\pm} K^{\mp} $} &
 \multirow{2}{*}{$B_{s}^{0}(\bar{B}_{s}^{0})\rightarrow D^{\pm} \pi^{\mp} $} &
 \multirow{2}{*}{$B^{0}(\bar{B}^{0})\rightarrow D^{\pm} \pi^{\mp} $ }\\
 &&&\\
 \hline
\vspace{0.2cm}
\multirow{1}{*}{$C_f$ } &
\multirow{1}{*}{$-0.71^{+0.02+0.08+0.01}_{-0.02-0.05-0.01}$} &
\multirow{1}{*} {$-0.84^{+0.01+0.02+0.00}_{-0.01-0.01-0.01}$}&
\multirow{1}{*}{$-1.00\pm0.00\pm0.00\pm0.00$}  \\
\vspace{0.2cm}
\multirow{1}{*}{$S_f$ } &
\multirow{1}{*}{$-0.63^{+0.02+0.05+0.11}_{-0.02-0.07-0.06}$} &
\multirow{1}{*} {$-0.39^{+0.01+0.01+0.12}_{-0.01-0.01-0.08}$}&
\multirow{1}{*}{$0.035^{+0.002+0.005+0.002}_{-0.002-0.003-0.003}$} \\
\vspace{0.2cm}
\multirow{1}{*}{$A_f^{\Delta\Gamma}$ } &
\multirow{1}{*}{$-0.32^{+0.01+0.02+0.17}_{-0.01-0.02-0.16}$} &
\multirow{1}{*}{$-0.36^{+0.02+0.03+0.12}_{-0.02-0.03-0.09}$} &
\multirow{1}{*}{$-0.011^{+0.001+0.003+0.008}_{-0.001-0.002-0.007}$} \\
\vspace{0.2cm}
\multirow{1}{*}{$S_{\bar{f}}$ } &
\multirow{1}{*}{$-0.65^{+0.02+0.05+0.10}_{-0.02-0.06-0.05}$} &
\multirow{1}{*} {$-0.53^{+0.02+0.02+0.04}_{-0.02-0.03-0.01}$}&
\multirow{1}{*}{$0.034^{+0.002+0.006+0.003}_{-0.002-0.003-0.004}$}  \\
\vspace{0.4cm}
\multirow{1}{*}{$A_{\bar{f}}^{\Delta\Gamma}$ } &
\multirow{1}{*}{$-0.27^{+0.01+0.03+0.18}_{-0.01-0.04-0.17}$} &
\multirow{1}{*} {$-0.059^{+0.010+0.008+0.14}_{-0.010-0.021-0.14}$}&
\multirow{1}{*}{$-0.015^{+0.001+0.001+0.008}_{-0.001-0.001-0.006}$}\\
\hline\hline
\end{tabular}\label{t2}
\end{table}

By using the PQCD approach introduced in the above section, we can get the numerical results of branching ratios for the considered six decay channels, which are listed in Table \ref{t1}. According to the definitions in Eq.(\ref{equ:cpob}), the numerical results of $CP$ violation parameters are shown in Table \ref{t2}. In our theoretical calculations, we estimate three kinds of theoretical uncertainties. The first error comes from the hadronic parameters including the decay constants and shape parameters in wave functions of the $B_{(s)}$ and $D$ mesons, which are given in Sec. \ref{sec:pqcd}. The Second one comes from the higher order  perturbative QCD corrections containing the uncertainty of $\Lambda_{QCD}=0.25\pm0.05$ GeV and the choice of the factorization scales. The third kind of error is caused by the uncertainties of the CKM matrix elements and the CKM angles $\gamma$ and $\beta_{(s)}$. The CKM angle $\gamma$ is an input parameter in our paper that was adopted as $\gamma=(68_{-11}^{+10})^{\circ}$\cite{Beringer:1900zz}.

For the theoretical results of branching ratios, the hadronic inputs contribute the largest uncertainty and the CKM elements contribute little. The CKM angles have no influence on the branching ratios which are proportional to the square of amplitudes. In contradiction with the branching ratios, the largest uncertainty of the $CP$ violation parameters comes from the CKM angles which are weak phase. The sensitivity to the CKM angles makes the measurement of these five $CP$ violation parameters a good way to extract the angle $\gamma$. The uncertainties of hadronic inputs have little impact on the results of $CP$ violation parameters because they provide little contribution to the strong phase difference $\Delta$ . This fact makes the measurement of $CP$ violation parameters more reliable because there is little influence from the large uncertainties of hadronic inputs.

From Table \ref{t1}, we find that our numerical results of branching ratios are consistent with the experimental data. For example the combined branching ratio of decay channels $B_{s}^{0}\rightarrow D_{s}^{\pm} K^{\mp}$ from our calculation is $(1.83^{+0.89+0.42+0.09}_{-0.63-0.45-0.05})\times 10^{-4}$, which agrees well with the LHCb experiment's result\cite{Aaij:2012zz}. We should point out that there is a little difference of the definition of the branching ratio between experiment and theory. The branching ratios of $B_s$ decays are defined as time-integrated untagged rates by experimenters, while for theorists the branching ratios correspond to the untagged rate at time $t=0$. However the difference is quite small here so we neglect it\cite{DeBruyn:2012wj}. For the decay channels $B_{s}^{0}\rightarrow D^{\pm} \pi^{\mp} $, they can decay only via W exchange diagrams in the Standard Model. Because these pure annihilation type decays are power suppressed, the branching ratios are quite small. The branching ratio of $B^{0}\rightarrow D^{+} \pi^{-} $ is especially smaller than that of  $B^{0}\rightarrow D^{-} \pi^{+} $ because of the CKM suppression.

Our results of five $CP$ violation parameters including $C_f$, $S_f$, $A_f^{\Delta\Gamma}$, $S_{\bar{f}}$ and $A_{\bar{f}}^{\Delta\Gamma}$ are listed in Table \ref{t2}. However, there are few experimental measurements of CP violation parameters for these decays. For $B^{0}\rightarrow D^{\pm} \pi^{\mp} $ decays, there is another set of $CP$ parameters called $a$ and $c$ in experiment and the experimental results are $a=-0.030\pm 0.017$ and $c=-0.022\pm 0.021$\cite{Amhis:2012bh}. We can translate our results to the above ones by using the relations $a=-(S_f+S_{\bar{f}})/2$ and $c=-(S_f-S_{\bar{f}})/2$ to get $a=-0.035^{+0.002+0.002+0.003}_{-0.002-0.004-0.002}$ and $c=-0.001^{+0.002+0.004+0.002}_{-0.002-0.003-0.002}$. Our results of $a$ and $c$ are consistent with the experimental averages. The value of $c$ is near zero because of a very small strong phase difference $\Delta$. The first measurement of the $CP$ violation parameters in $B_{s}^{0}\rightarrow D_{s}^{\pm} K^{\mp} $ has recently been made \cite{Blusk:2012v3}. However the experimental errors are quite large and we expect more precise results in the future experiments.

\begin{figure}[]
\begin{center}
\includegraphics{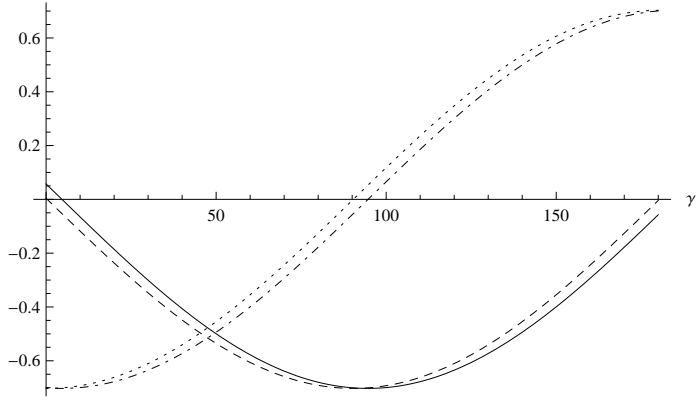}
\caption{$CP$ violation parameters of $B_{s}^{0}(\bar{B}_{s}^{0})\rightarrow D_{s}^{\pm} K^{\mp}$: $S_f$(solid line), $A_f^{\Delta\Gamma}$(dot-dashed line), $S_{\bar{f}}$(dashed line) and $A_{\bar{f}}^{\Delta\Gamma}$(dotted line) as functions of the CKM angle $\gamma$.  }
\label{fig:gam}
\end{center}
\end{figure}

The four $CP$ violation parameters $S_f$, $A_f^{\Delta\Gamma}$, $S_{\bar{f}}$ and $A_{\bar{f}}^{\Delta\Gamma}$ are $\gamma$ related. The relationship between $CP$ parameters and $\gamma$ in $B_{s}^{0}\rightarrow D_{s}^{\pm} K^{\mp}$ decays are shown in Fig.\ref{fig:gam}, with $\gamma$ ranging from $0$ to $180$ degree. The curves are trigonometric functions due to the definition in Eq.(\ref{equ:cpob}) and Eq.(\ref{equ:twol}). If we measure these parameters from experiments, we can extract the CKM angle $\gamma$ and strong phase by using Fig.\ref{fig:gam}.

Finding modes where the strong phase difference $\Delta$ equals to zero is important because $\sin(\gamma+2\beta)$ can be extracted without any ambiguity if $\Delta$ is negligible\cite{Aleksan:1991nh}. The decays $B_{s}^{0}(\bar{B}_{s}^{0})\rightarrow D_{s}^{\pm} K^{\mp}$ and  $B^{0}(\bar{B}^{0})\rightarrow D^{\pm} \pi^{\mp} $ are such ideal modes with $\Delta \approx 0$. The dominant contributions of these two decay modes come from the factorizable emission diagrams in Fig.\ref{fig:feyd}(a) and \ref{fig:feyd}(b) which contribute no strong phase. Although other six diagrams shown in Fig. \ref{fig:feyd}(c)-\ref{fig:feyd}(h) contribute strong phase, the amplitudes of them are quite small. Therefore the strong phase difference $\Delta$ is close to zero. From Eq.(\ref{equ:twol}) the phase difference between $\lambda_f$ and $\lambda_{\bar{f}}$ is $2 \Delta$, so a very small $\Delta$ means the difference between $S_f$ and $S_{\bar{f}}$ and the difference between $A_f^{\Delta\Gamma}$ and $A_{\bar{f}}^{\Delta\Gamma}$ are very small, which is consistent with the numerical results in Table \ref{t2} and Fig.\ref{fig:gam}. For the decay mode $B_{s}^{0}(\bar{B}_{s}^{0})\rightarrow D^{\pm} \pi^{\mp} $, they can decay only via annihilation type  diagrams which provide a large strong phase in the PQCD approach. Therefore the strong phase difference $\Delta$ is not close to zero, and there is no reason to ask for the small difference between $S_f$ and $S_{\bar{f}}$ and between $A_f^{\Delta\Gamma}$ and $A_{\bar{f}}^{\Delta\Gamma}$.

An ideal value of $\lambda_f$ should be at the order of $1$, because if $\lambda_f$ is too large or too small, $C_f$ will be close to $+1$ or $-1$ according to Eq. (\ref{equ:cpob}), which requires a high experimental resolution in order to derive $\lambda_f$ from $C_f$. Another reason is that a too large or too small $\lambda_f$ will make the $CP$ violation parameters $S_f$ and $A_{f}^{\Delta\Gamma}$ too small to be measured. The decay modes $B_{s}^{0}(\bar{B}_{s}^{0})\rightarrow D_{s}^{\pm} K^{\mp}$ and $B_{s}^{0}(\bar{B}_{s}^{0})\rightarrow D^{\pm} \pi^{\mp} $ have such proper $\lambda_f$, because from the CKM elements we can roughly estimate the order of $\lambda_f \sim |V_{us}^* V_{cb}/V_{ub}^*V_{cs}|\sim 1$. However, for $B^{0}(\bar{B}^{0})\rightarrow D^{\pm} \pi^{\mp} $ decays, we just take $d$ quark instead of $s$ quark in $B_{s}^{0}(\bar{B}_{s}^{0})\rightarrow D_{s}^{\pm} K^{\mp} $ decays. So $\lambda_f$ is proportional to $ |V_{ud}^* V_{cb}/V_{ub}^*V_{cd}|\sim 1/\lambda^2$ to yield $\lambda_f=54$, which is quite large and hard to be  measured in experiment.

To overcome the shortcoming of $B^{0}(\bar{B}^{0})\rightarrow D^{\pm} \pi^{\mp}$ decays, the large $\lambda_f$, we also explore $B^{0}(\bar{B}^{0})\rightarrow D^{\pm} a_2^{\mp}$ decays, with a tensor meson $a_2(1320)$ instead of the pseudoscalar meson $\pi$\cite{Wang:2012jba}.
For B to tensor decays, there is a special property that the factorizable amplitude with a tensor meson emitted vanishes because of $\langle0|j^{\mu}|T\rangle=0$, where $j^{\mu}$ is the $(V\pm A)$ current or $(S\pm P)$ density\cite{Kim:2001sha,Kim:2001py,Cheng:2010hn,Cheng:2010yd,Zou:2012sx}. Although the amplitude of $B^{0}\rightarrow D^{+}a_2^{-}$ is suppressed by the CKM matrix elements, the amplitude of $\bar{B}^{0}\rightarrow D^{+}a_2^{-}$ is also suppressed because there is no factorizable emission diagrams with $a_2^{-}$ emitted. Therefore we expect $\lambda_f$ of $B^{0}(\bar{B}^{0})\rightarrow D^{\pm} a_2^{\mp}$ decays may not be so big as the $B^{0}(\bar{B}^{0})\rightarrow D^{\pm} \pi^{\mp}$ decays. In fact we get $\lambda_f=16$ from our calculation, smaller than that of $B^{0}(\bar{B}^{0})\rightarrow D^{\pm} \pi^{\mp}$ but not small enough. The reason is that nonfactorizable emission diagrams and annihilation diagrams make a large contribution. The branching ratios and $CP$ violation parameters are listed below. This decay mode could be another choice to extract the CKM angle $\gamma$.
\begin{eqnarray}
Br(B^{0}\rightarrow D^{+}a_2^{-})&=&(1.75^{+0.61+0.30+0.16}_{-0.44-0.34-0.15})\times 10^{-6},\nonumber\\
Br(B^{0}\rightarrow D^{+}a_2^{-})&=&(4.35^{+1.30+1.34+0.23}_{-1.01-1.43-0.11})\times 10^{-4},
\end{eqnarray}
\begin{eqnarray}
C_f&=&-0.99^{+0.00+0.01+0.00}_{-0.00-0.00-0.00},\nonumber\\
S_f&=&-0.12\pm0.01\pm0.02\pm0.01,\nonumber\\
A_f^{\Delta \Gamma}&=&-0.047^{+0.005+0.030+0.025}_{-0.005-0.044-0.024},\\
S_{\bar{f}}&=&-0.057^{+0.001+0.015+0.024}_{-0.001-0.006-0.023},\nonumber\\
A_{\bar{f}}^{\Delta \Gamma}&=&0.11^{+0.01+0.05+0.01}_{-0.01-0.03-0.02}.\nonumber
\end{eqnarray}

\section{SUMMARY}\label{sec:sum}
In this paper, we investigate the time-dependent $CP$ violations of $B_{s}^{0}(\bar{B}_{s}^{0})\rightarrow D_{s}^{\pm} K^{\mp} $, $B_{s}^{0}(\bar{B}_{s}^{0})\rightarrow D^{\pm} \pi^{\mp} $ and  $B^{0}(\bar{B}^{0})\rightarrow D^{\pm} \pi^{\mp} $ decays within the framework of the PQCD approach. We predicted branching ratios and $CP$ violation parameters, providing theoretical expectation for future experiment measurements. The branching ratios of $B_{s}^{0}(\bar{B}_{s}^{0})\rightarrow D_{s}^{\pm} K^{\mp} $ and $B^{0}(\bar{B}^{0})\rightarrow D^{\pm} \pi^{\mp} $ and the CP asymmetry of $B^{0}(\bar{B}^{0})\rightarrow D^{\pm} \pi^{\mp} $ calculated are consistent with the experimental data. From the above discussions, we can see that $B_{s}^{0}(\bar{B}_{s}^{0})\rightarrow D_{s}^{\pm} K^{\mp} $ is the most favorable decay mode to extract $\gamma$, because it has
a large branching ratio and a proper $\lambda_f$.

\section*{Acknowledgment}

We are very grateful to Fu-Sheng Yu  and Qin Qin for helpful discussions. This work is partially supported
by National Science Foundation of China under the Grant No.11075168,11228512 and 11235005.

\bibliography{timedependentCP}

\end{document}